\documentstyle[aps,prc,epsfig,multicol]{revtex}
\newcommand{\he}{{$^3$He}}
\newcommand{\etal}{{\it et al.}}

\newcommand{\prev}{Phys.~Rev. }
\newcommand{\nim}{Nucl.~Instr.~Meth. }

\begin{document}
\title{Stimulated Emission of Radiation in a Nuclear Fusion
  Reaction\footnote[1]{Submitted to Phys. Rev. C; Copyright 1999 by
    The American Physical Society}}
\author{Michael D\"uren}

\address{ Physikalisches Institut, Universit\"at Erlangen-N\"urnberg,
  91058 Erlangen, Germany\\
  {\small E-mail: {\tt Michael.Dueren@DESY.DE} }}
\date{\today}
\maketitle
\begin{abstract}
  This letter claims that process of stimulated emission of radiation
  can be used to induce a fusion reaction in a HD molecule to produce
  \he. An experimental set-up for this reaction is presented. It is
  proposed to study the technical potential of this reaction as an
  energy amplifier.

\end{abstract}

\begin{multicols}{2}[]
  Standard techniques for nuclear fusion require high temperature and
  pressure to overcome the Coulomb barrier of the repulsing forces
  between the reaction partners. Here, a completely different approach
  to fusion is discussed. The electromagnetic, exothermic fusion
  reaction
\begin{equation}
  \label{eq:fusi}
  \mbox{p} +\mbox{d}\to \mbox{$^3$He}+\gamma \quad  \mbox{($Q=$ 5.49 MeV)}
\end{equation}
is well known as part of the solar cycle~\cite{solar1,solar2}. When
the initial state is replaced by a HD molecule, the reaction can be
treated as an electromagnetic transition between two quantum levels,
the HD and the \he\ state.  In ana\-logy to a laser, three types of
transitions can be distinguished~\cite{stimul1,stimul2}, the
spontaneous emission
\begin{equation}
  \label{eq:spont}
  \mbox{HD}\to \mbox{$^3$He}+\gamma \, ,
\end{equation}
the absorption
\begin{equation}
  \label{eq:abso}
\gamma+  \mbox{$^3$He} \to\mbox{HD} \, ,
\end{equation}
and the stimulated emission
\begin{equation}
  \label{eq:stimu}
\gamma+  \mbox{HD}\to \mbox{$^3$He}+2\gamma \, .
\end{equation}

The spontaneous emission happens at finite time and is given by the
quantum mechanical tunneling of the p,d nuclei across the Coulomb
barrier. The repulsing forces between the nuclei are many orders of
magnitude larger than the molecular binding forces. Therefore the
spontaneous emission is largely suppressed to a negligible probability
and is only of academic interest.

The absorption process $\gamma+ \mbox{$^3$He} \to\mbox{p}+\mbox{d}$ is
well known in literature, is called photodisintegration and has been
measured previously~\cite{discross,discross1}. Its cross section has a
maximum of about 0.8~mb at a photon energy of 12~MeV.  Neglecting the
binding electrons, the process from Eq.~(\ref{eq:abso}) is given by
the photodisintegration cross section close to threshold.

The process of interest is the stimulated emission. In this case the
fusion process is induced by the electromagnetic field of a 5.49~MeV
initial photon. This field has more energy than needed to overcome the
Coulomb barrier of the p,d-system. A proton energy below 0.3~MeV has
been shown to be sufficient to induce a fusion reaction
D(p,$\gamma$)\he\ with a cross section of 1~$\mu$b~\cite{barrier}.
The stimulated emission by a 5.49~MeV photon requires no tunneling,
with the effect, that the suppression factor which is present in the
spontaneous process does not apply here. By formally adding a photon
on both sides in Eq.~(\ref{eq:abso}): $2\gamma+ \mbox{$^3$He} \to
\mbox{HD}+\gamma$ the analogy between the processes of absorption and
of stimulated emission is demonstrated.

From energy-momentum conservation follows that due to the recoil of
the $^3$He nucleus the energy of the emitted photon is $\Delta E_{\rm
  red}=5.4 $~keV below the $Q$-value of the reaction (1). A
characteristic property of the process of stimulated emission is that
the two outgoing photons are coherent, meaning that their wave
functions have the same phase and that the two photons occupy an
identical element in the phase space.  To match this requirement, the
incoming photon has to have the same energy, i.e. it also has to have
an energy which is 5.4~keV below the $Q$-value of the reaction (1).
This means that the cross section of the stimulated emission as a
function of the photon energy is a $\delta$-function at $E=Q-\Delta
E_{\rm red}$. Due to the long lifetime of the HD
molecules, there is no measurable intrinsic width of this resonance.
The effective width $\delta E_{\rm eff}$ of the resonance is
determined by the Doppler motion of the HD molecule and by the
electronic excitations of the $^3$He final state.  Nuclear excitations
of the final state are not discussed here. They lead to additional
distinct resonances at lower photon energy.  The Doppler motion
depends on the temperature of the HD target and is small, especially
for a cryogenic target. As there is no special mechanism to accelerate
the shell electrons, it is natural to assume that the ionization
energy of $^3$He is a good approximation for the energy uncertainty of
the final state. The ionization energy is 79 eV~\cite{molecule}.  This
leads to an extremely peaked resonance with a relative width of the
order of $\delta E_{\rm eff}/E\approx 1.4\cdot 10^{-5}$.

The actual size of the cross section at the peak is hard to
estimate. A general thermodynamic consideration as in
Ref.~\cite{stimul1} leads to the conclusion that the 
transition probabilities for absorption and for stimulated emission
have to be identical in order to be compatible with the Planck law of
black body radiation. Assuming that this consideration can be applied
to our process, the cross section has to have the same order of
magnitude as the absorption cross section from Eq.~(\ref{eq:abso}).

In trying to understand the process of stimulated emission on the
microscopic level, the following considerations might be of interest.
The Coulomb barrier of the p,d-system keeps the wave functions of the
two nuclei separated to molecular distances, i.e.~to a distance which
corresponds to a fraction of the average separation of the nuclei in
the HD molecule which is $0.74\cdot 10^{-10}$m~\cite{molecule}.  Due
to the large distance, an incoming photon will usually react
incoherently with one of the nuclei, and the chance that the struck
nucleus will react in a second step with the other nucleus is unlikely
and defined by the fusion cross section which is in the $\mu$b
range~\cite{barrier}.

The process of stimulated emission requires a coherent reaction of p
and d. The phrase `coherent reaction' in this context means that the
photon reacts with the total quantum system of the HD molecule instead
with the individual nuclei. As the p and d wave functions do not
overlap, it is required that the photon has a spatial extension which
is large is enough to overlap with both wave functions. The
longitudinal extension of an individual photon, i.e.~of its wave
package, is defined by the coherence length of the individual photon.
Here, photons with a coherence length of molecular dimensions are
required.  From the Heisenberg uncertainty principle follows that the
photons which are produced by stimulated emission fulfill the
requirement, as their energy is determined with a resolution of the
order of $\sim 80$~eV and therefore their longitudinal extension has
to be larger than $\hbar c\cdot (80 $~eV$)^{-1}=0.25\cdot 10^{-10}$m.
This means that photons from HD fusion have a coherence length which
is sufficient to stimulate further fusion reactions.

A way to visualize the coherent process in a semi-classical way is to
imagine that the electromagnetic field of such a photon raises for a
short moment the energy of the two nuclei above the Coulomb barrier.
Once the energy of the nuclei is above the barrier, the wave functions
of the nuclei can expand, fluctuate and occasionally overlap.  Even if
the overlap happens only for a short moment, nuclear forces will
immediately lead to a collapse of the wave functions into the
low-energy $^3$He state. There is no quantum number violation which
prohibits the stimulated fusion reaction and the author is not aware
of any other mechanism which suppresses this process beyond the
typical electromagnetic cross sections. A simple order of magnitude
estimate of the cross section is given by the magnetic dipole photon
absorption cross section~\cite{blatt}
\begin{equation}
  \label{eq:dipole}
  \sigma \approx 4E\cdot 0.48\mbox{ mb/MeV }=10.5 \mbox{ mb}.
\end{equation}
The factor 4 comes from adding the amplitudes for the two nuclei
coherently.

As in a laser device, the reaction of stimulated emission can be used
as photon amplifier.  A necessary criterion for a laser to work is the
population inversion of the laser transition. In our case the
population inversion corresponds to the case that there are more
particles in the HD state than in the \he\ state.  This criterion is
trivially achieved in any HD target. As in an ordinary laser the
stimulated emission works iteratively, i.e.~the emitted photons induce
further processes of stimulated emission while they pass through the
laser material. The final intensity grows approximately exponentially
with the length of the target.

To initiate the reaction, a photon from an external source is
required. A usual external source typically does not fulfill the
coherence length requirement, as e.g.~photons from typical nuclear
reactions have nuclear dimensions, and therefore it is unlikely that
those photons will induce stimulated emission. Instead, photons from
coherent bremsstrahlung are emitted in a collective process where
several nuclei in a crystal are involved~\cite{cohbrems1,cohbrems2}.
Therefore, the individual photons from coherent bremsstrahlung can
have a coherence length which has molecular dimensions and therefore
might be able to induce the process of stimulated emission in a HD
target.

Up to now, no background processes have been taken into account.  The
HD target serves only as photon amplifier when the cross section of
stimulated fusion is larger than the cross section of other
electromagnetic processes at this energy like e.g.~pair production or
bremsstrahlung. The total cross section of background processes for an
unpolarized beam and target is 160 mb/molecule, as calculated from
Ref.~\cite{pdg}. If the rough cross section estimate of
Eq.~(\ref{eq:dipole}) is correct, the background dominates and the
photon beam is exponentially attenuated instead of amplified.
Nevertheless, nuclear binding energy is released when an external beam
hits the HD target.  To improve the fusion to background ratio, it can
be considered to spin polarize the photon beam and the HD target.

\begin{figure}[tph]
\centerline{
\epsfxsize 16cm
\epsfbox{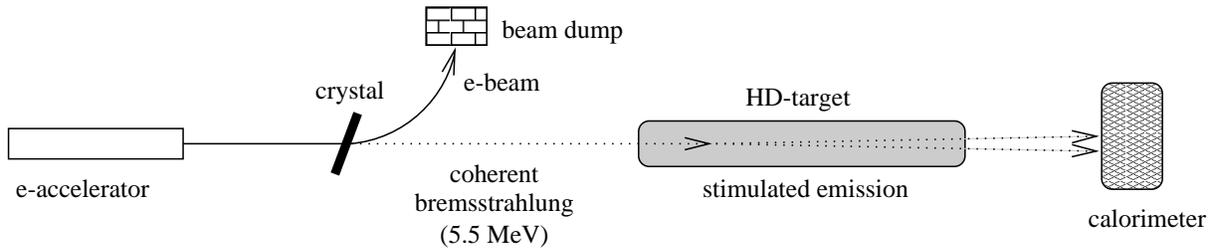}
  }
\smallskip
\caption{\small A possible set-up for an experiment to study the
  stimulated emission of photons in the fusion process $ \mbox{p}
  +\mbox{d}\to \mbox{$^3$He}+\gamma$. A photon beam hits a (liquid) HD
  target and induces fusion processes. The photons are detected in a
  calorimeter.
  \label{sketch}}
\end{figure}

A possible scenario for an experiment to study the fusion process and
the energy amplification is sketched in Fig.~\ref{sketch}. A beam with
5.49~MeV photons that have a large coherence length is generated by
coherent bremsstrahlung of an electron beam incident on a crystal. The
photons hit a liquid or solid HD target. A calorimeter measures the
energy of individual photons. If the process of stimulated emission
exists, it will show up as coincidence of two or more photons in the
same calorimeter element which add up their deposited energy. A peak
at $2\cdot 5.49$~MeV$=10.98$~MeV will appear in the energy spectrum of
the calorimeter. A second target which contains a mixture of H$_2$ and
D$_2$ should not show the effect. A photon beam with a high duty cycle
is preferred for this experiment to reduce random coincidences.  Once
the effect has been demonstrated, it can be considered to repeat the
experiment with a spin-polarized target and beam to enhance the signal
to background ratio.  Experimentally, the techniques of controlling
the polarization of HD targets~\cite{hdpol} and coherent
bremsstrahlung beams~\cite{beampol} are well understood.

To summarize, it is claimed that the process of stimulated emission of
radiation can be used to induce a fusion reaction in HD to produce
\he. Thermodynamic considerations suggest that the cross section of
stimulated fusion is related to the cross section of
photodisintegration of \he\ at threshold. From energy momentum
conservation in connection with the almost infinite lifetime of
initial and final state follows that the cross section as function of
photon energy is a sharp resonance. If a set-up can be found where
this or a similar reaction exceeds other competing electromagnetic
processes, the stimulated emission can be technically used as an
energy amplifier which is fueled by nuclear energy.  First order of
magnitude estimates indicate an insufficient size of the cross section
with the consequence that this process can not be used as photon
amplifier. However, it is estimated that the cross section is in a
range where the process can be experimentally demonstrated. This
letter is meant as an invitation to calculate and measure the cross
section of stimulated fusion or to prove that this process does not
exist.


\end{multicols}

\begin{references}
\bibitem{solar1} H. A. Bethe and C. L. Critchfield, \prev {\bf 54} (1938) 248.
\bibitem{solar2} E. E. Salpeter, Ann.~Rev.~Nucl.~Sci., Vol.~2 (1953) 41.
\bibitem{molecule} D. R. Lide (ed.), {\em CRC Handbook of Chemistry
  and Physics}, CRC Press Inc., Boca Raton, Florida (1998).
\bibitem{stimul1} A. Einstein, Phys.~Z.~{\bf 18} (1917) 121;
  English translation in D. ter Haar, {\em The Old Quantum Theory},
  Pergamon, Oxford (1967).
\bibitem{stimul2}  R. P. Feynman, {\em Quantum Electrodynamics},
  Benjamin/ Cummings Publishing Company, Menlo Park, California, USA (1961). 
\bibitem{discross} A. N. Gorbunov and A. T. Varfolomeev, \pl {\bf 11}
  (1964) 137. 
\bibitem{discross1} G. Ticcioni, S.N. Gardiner, J.L. Matthews and R.O. Owens,
  \pl B {\bf 46} (1973) 369. 
\bibitem{barrier} G. M. Griffiths, E. A. Larson and L. P. Robertson,
  Can.~J.~Phys.~{\bf 40} (1962) 402.
\bibitem{blatt} J.M. Blatt and V.F. Weisskopf, {\em Theoretical
    Nuclear Physics} p. 654, Wiley, New York (1952). 
\bibitem{pdg}Particle Data Group, Eur. Phys. J. C {\bf 3}, (1998) 1\\
 and  {\tt http://physics.nist.gov/PhysRefData}.
\bibitem{cohbrems1}H. \"Uberall, \prev {\bf 103} (1956) 1055.
\bibitem{cohbrems2}G. Diambrini Palazzi, \rmp {\bf 40} (1968) 611.
\bibitem{hdpol} A. Honig {\em et al.}, Proc.~of the ``12th
  International Symposium on High-Energy Spin Physics,'' edited by
  C.W. de Jager \etal, Amsterdam, The Netherlands, World Scientific
  (1997) 365.
\bibitem{beampol} D. Lohmann {\em et al.}, \nim A {\bf 343} (1994) 494.
\end{references}
\end{document}